\documentclass[letterpaper,12pt,tightenlines,notitlepage,superscriptaddress,nofootinbib,showpacs]{revtex4-1}
\usepackage{amsmath,amssymb}
\usepackage[pdftex]{graphicx}
\usepackage{mathptmx}
\usepackage{color}
\usepackage{bm,wasysym,mathtools,xfrac}
\usepackage[pdftex,unicode,colorlinks]{hyperref}
\usepackage[utf8x]{inputenc}

\newcommand{\eg}{{\it e.g.}\,}
\newcommand{\etc}{{\it etc}}
\newcommand{\hc}{\mathrm{h.c.}}
\newcommand{\CC}{\mathrm{C.C.}}
\renewcommand{\Re}{\mathop{\rm Re}\nolimits}
\renewcommand{\Im}{\mathop{\rm Im}\nolimits}
\newcommand{\sign}{\mathop{\rm sign}\nolimits}
\newcommand{\mean}[1]{\langle#1\rangle}
\newcommand{\Mean}[1]{\left\langle#1\right\rangle}
\newcommand{\partd}[2]{\dfrac{\partial#1}{\partial#2}}
\newcommand{\svector}[2]{\begin{pmatrix}#1 \\ #2 \end{pmatrix}}
\newcommand{\smatrix}[4]{\begin{pmatrix}#1 & #2 \\ #3 & #4\end{pmatrix}}

\begin{document}

\title{Generalized analysis of quantum noise and dynamic back-action in signal-recycled Michelson-type laser interferometers}

\author{Farid Ya.~Khalili}
\email[]{Corresponding author: khalili@phys.msu.ru}
\affiliation{M.V.Lomonosov Moscow State University, Faculty of Physics, Moscow 119991, Russia}

\author{Sergey P. Tarabrin}
\affiliation{Institut f\"{u}r Gravitationsphysik, Leibniz Universit\"{a}t Hannover and Max-Planck Institut f\"{u}r Gravitationsphysik (Albert-Einstein Institut),\\ Callinstra\ss{}e 38, D-30167 Hannover, Germany}
\affiliation{Institut f\"ur Theoretische Physik, Leibniz Universit\"at Hannover, Appelstra\ss{}e 2, D-30167 Hannover, Germany}

\author{Roman Schnabel}
\affiliation{Institut f\"ur Laserphysik and Zentrum f\"ur Optische Quantentechnologien, Universit\"{a}t Hamburg, D-22761 Hamburg, Germany}

\author{Klemens Hammerer}
\affiliation{Institut f\"{u}r Gravitationsphysik, Leibniz Universit\"{a}t Hannover and Max-Planck Institut f\"{u}r Gravitationsphysik (Albert-Einstein Institut),\\ Callinstra\ss{}e 38, D-30167 Hannover, Germany}
\affiliation{Institut f\"ur Theoretische Physik, Leibniz Universit\"at Hannover, Appelstra\ss{}e 2, D-30167 Hannover, Germany}


\begin{abstract}
  We analyze the radiation pressure induced interaction of mirror motion and light fields in Michelson-type interferometers used for the detection of gravitational waves and for fundamental research in table-top quantum optomechanical experiments, focusing on the asymmetric regime with a (slightly) unbalanced beamsplitter and a (small) offset from the dark port. This regime, as it was shown recently, provides new interesting features, in particular a stable optical spring and optical cooling on cavity resonance.
  
   We show that generally the nature of optomechanical coupling in Michelson-type interferometers does not fit into the standard dispersive/dissipative dichotomy. In particular, a symmetric Michelson interferometer with signal-recycling but without power-recycling cavity is characterized by a purely dissipative optomechanical coupling; only in the presence of asymmetry, additional dispersive coupling arises.  In gravitational waves detectors possessing signal- and power-recycling cavities, yet another, ``coherent'' type of optomechanical coupling takes place. 
   
   We develop here a generalized framework for the analysis of asymmetric Michelson-type interferometers which also covers the possibility of the injection of carrier light into both ports of the interferometer. Using this framework, we analyze in depth the ``anomalous'' features of the Michelson-Sagnac interferometer which where discussed and observed experimentally in previous works \cite{Xuereb_PRL_107_213604_2011, 13a1TaKhKaScHa, 14a1SaKaNiTaKhHaSc}. 
\end{abstract}

\maketitle

\section{Introduction}

The Michelson interferometer was first used in 1887 in the famous experiment by A.~Michelson and E.~Morley \cite{Michelson1887}. Since then, it became a standard tool being routinely employed in high-precision optical measurements. Currently, the most conspicuous devices based on the Michelson interferometer topology are gravitational-wave (GW) detectors, like LIGO \cite{LIGOsite, CQG_32_7_074001_2015}, VIRGO \cite{AdvVIRGOsite, Accadia2012}, and GEO-600 \cite{GEOsite, Grote2010}, which have arm lengths varying form several hundreds of meters to several kilometers.

\begin{figure}
  \includegraphics{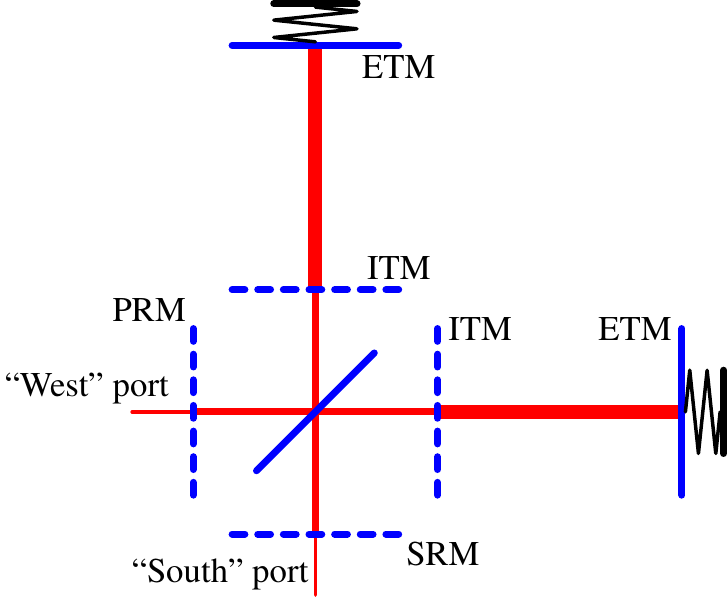}
  \caption{The dual-recycled Michelson/Fabry-Perot topology of the modern laser GW detectors. PRM: the power recycling mirror; SRM: the signal recycling mirror; ITM: the input test mass; ETM: the end test masses. The optional mirrors are shown by dashed lines (in the real GW detectors, either ITMs, or PRM and/or SRM can be absent).}\label{fig:dual_rec_Michelson}
\end{figure}

The typical optical layout of GW detectors is shown in Fig.\ \ref{fig:dual_rec_Michelson}. In addition to the end mirrors (the end test masses, ETMs), it could include up to four additional ones. Two of them (the input test masses, ITMs), form, together with the  ETMs, two Fabry-Perot arm cavities, which increase the light's storage time for improving the interferometer's signal response. Two so-called recycling mirrors, the power- and the signal-recycling mirror (PRM and SRM) allow to independently tune the bandwidths and the detunings of its two optical modes, the common and the differential ones  \cite{Vinet_PRD_38_433_1988, Meers1988}. Detuning of the SR mirror can also result in a sensitivity improvement via the so-called `optical spring' \cite{Buonanno2002}. Since it is dynamically unstable, also schemes exploiting two bright light fields were researched in order to provide a stable optical spring \cite{Corbitt2007,Rehbein2008}.

\begin{figure}
  \includegraphics{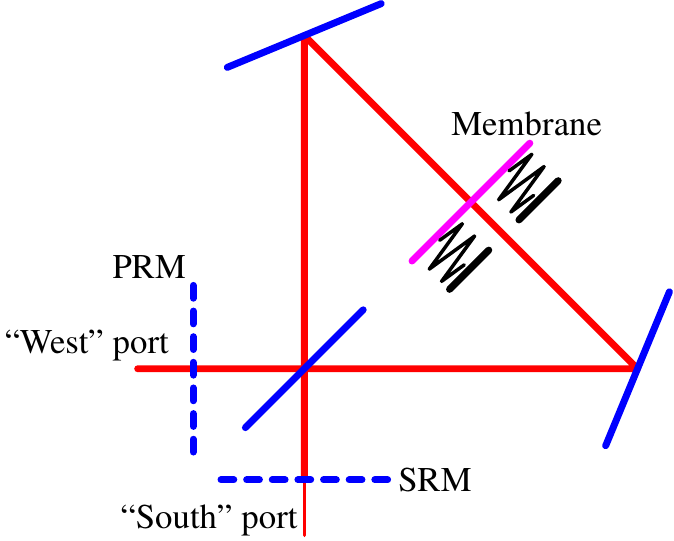}
  \caption{The Michelson-Sagnac interferometer. PRM: the power recycling mirror; SRM: the signal recycling mirror. The optional mirrors are shown by dashed lines.}\label{fig:Michelson-Sagnac}
\end{figure}

Several years ago, the Michelson interferometer topology was adopted also for table top quantum optomechanical experiments, with partly translucent silicon-nitride membranes playing the role of the test mass \cite{09a1YaFrWeGoDaScSoDa}. These membranes have very small masses ($m\lesssim100\,{\rm ng}$) and low optical and mechanical losses and provide a suitable platform for quantum optomechanical experiments \cite{Thompson_Nature_452_72_2008}. They have, however, a relatively low reflectivity, which does not allow to use them as end mirrors in high-finesse optical resonators. Instead, the {\it Michelson-Sagnac} topology was proposed in Ref.~\cite{09a1YaFrWeGoDaScSoDa}, see Fig.\,\ref{fig:Michelson-Sagnac}. It can be viewed as a derivative of the dual-recycled (signal- and power-recycled) Michelson topology of laser GW detectors. By folding the Michelson arms towards each other, light that is transmitted through the membrane does not leave the interferometer, and the membrane takes the role of the end mirror of both Michelson arms. 
In turn, the Michelson interferometer can be treated as a special case of the Michelson-Sagnac interferometer, when setting the membrane transmissivity equal to zero. The general theory of the dual-recycled Michelson-Sagnac interferometer presented in this article can be applied to all {\it Michelson-type interferometers}  -- the Michelson-Sagnac, the pure Michelson and the Michelson-Fabry-Perot interferometer.

The standard and well-explored regime of these interferometers assumes a balanced beam splitter, interferometer arms of identical length and optical loss as well as an operation at (or very close to) a dark fringe. This is what we call `\emph{the symmetric regime}'. 
A detailed analysis of the dual-recycled Michelson-Fabry-Perot interferometer in the symmetric regime was presented in \cite{Buonanno2003}. It was shown, that the complete interferometer can be mapped to a single Fabry-Perot cavity with effective parameters (the so-called scaling law theorem). 
Later the analysis was extended to the symmetric Michelson-Sagnac interferometer \cite{16a1DaKh}.

The first analysis in the \emph{asymmetric} regime of the Michelson-Sagnac interferometer was performed in \cite{Xuereb_PRL_107_213604_2011}. Here, it was in particular shown that optical ground state cooling is possible even outside good cavity regime \cite{Aspelmeyer_RMP_86_1391_2014}, which is due to a ``Fano resonance'' shape of the radiation pressure noise spectral density \cite{Elste_PRL_102_207209_2009, Weiss_Burder_Nunnen_2012}. In \cite{13a1TaKhKaScHa, 14a1VoVy}, the dynamic back action (that is, the optical spring features \cite{Buonanno2002}) of the asymmetric Michelson-Sagnac was analyzed and it was shown, that in contrast to the symmetric case, both the optical damping and the optical rigidity in an asymmetric Michelson-Sagnac interferometer could acquire a nonzero value on the optical resonance, and additional stability and instability regions exist on either side of the resonance. Later, this non-canonical behavior was demonstrated experimentally \cite{14a1SaKaNiTaKhHaSc}.

Here we present the generalized framework for the analysis of asymmetric cavity-enhanced Michelson-type interferometers that includes not only dynamical optomechanical back-action but also the light's quantum noise. In particular, we assume that both input/output ports of the interferometer can be pumped; this assumption simplifies the analysis of the interferometer and provides insights into the internal structure of the equations obtained in \cite{13a1TaKhKaScHa}. In Sec.\,\ref{sec:modes} we show that the character of the optomechanical coupling in  Michelson-type interferometers depends on whether one or two  recycling mirrors are present. In Sec.\,\ref{sec:calculations}, we analyze in detail the case of just one (signal-) recycling cavity, using the developed framework to explain the ``anomalous'' features of \cite{13a1TaKhKaScHa, 14a1SaKaNiTaKhHaSc}. In Sec.\,\ref{sec:cooling} we provide the optimization of optical cooling in Michelson-type interferometers. The notations used throughout this paper areis given in Table\,\ref{tab:notations}.

\begin{table}
  \begin{ruledtabular}
    \begin{tabular}{ll}
      Quantity & Description \\ \hline
      $c$                 & Speed of light \\
      $\hbar$             & Reduced Plank constant \\
      $\kappa_B$       & Boltzmann constant \\
      $\omega = ck$       & Any high (optical) frequency \\
      $\omega_p = ck_p$   & Optical pump frequency \\
      $\gamma$            & Optical half-bandwidth \\
      $\delta=\omega_p-\omega_o$ & Detuning of the pump from the optical resonance
      frequency \\
      $\Omega$            & Any low (mechanical) frequency; if appears together with $\omega$, then $\Omega=\omega-\omega_p$ \\
      $L_S = c\tau_S$     & Optical distance between the SRM and the symmetry position of the membrane \\
      $L_P = c\tau_S$     & The same for the PRM \\
      $R_{W,S}$           & Amplitude reflectivities of the power (W) and signal (S) recycling mirrors \\
      $T_{W,S}$           & Amplitude transmissivities of these mirrors \\
      $R_m = \cos\theta$  & Amplitude reflectivity of the membrane \\
      $T_m = \sin\theta$  & Amplitude transmissivity of the membrane \\
      $R = \cos(\pi/4-\epsilon)$  & Amplitude reflectivity of the beamsplitter \\
      $T = \sin(\pi/4-\epsilon)$  & Amplitude transmissivity of the beamsplitter \\
      $X = \varkappa/k_p$ & D.C. component of the membrane  displacement from the symmetry position\\
      $x$                 & A.C. component of the membrane  displacement from the symmetry position\\
      $\hc$               & Hermitian conjugate of the previous term \\
      $\CC$               & Caves-Schumaker conjugate \cite{Caves1985} of the previous term, see Eq.\,\eqref{def_CC}.
    \end{tabular}
  \end{ruledtabular}
  \caption{Main notations used in this paper.}\label{tab:notations}
\end{table}

\section{Optomechanical coupling in Michelson-type interferometers}\label{sec:modes}

In order to provide the starting point for our consideration below, let us start with the well-explored case of a single optical mode whose eigenfrequency depends on the position of the mechanical object. This type of the optomechanical coupling is known as the {\it dispersive} one. The Hamiltonian of this system can be presented in the standard form
\begin{equation}\label{H_cav}
  \hat{\mathcal{H}}
  = \hbar(\omega_o - g\hat{x})(\hat{{\rm e}}^\dagger\hat{{\rm e}} + \sfrac{1}{2})
    + \hat{\mathcal{H}}_m + \hat{\mathcal{H}}_{\rm rest} \,,
\end{equation}
where $\hat{{\rm e}}$ and $\hat{{\rm e}}^\dagger$ are the annihilation and creation operators of the intracavity field (we reserve the notation $\hat{{\rm a}}$ for the incident field), $\hat{x}$ is the mechanical coordinate, $\omega_o$ and $g$ are the optical eigenfrequency and the optomechanical coupling factor, $\mathcal{H}_m$ is a mechanical Hamiltonian and $\mathcal{H}_{\rm rest}$ is the Hamiltonian describing all other optical degrees of freedom, including the optical pump(s) and the optical losses. Note that the Fabry-Perot cavity treatment can be reduced to this lumped mode model, provided that one of its optical modes is selected by the strong classical pump with the frequency $\omega_p$ close to this mode eigenfrequency.

Following Sec.\,III of the review paper \cite{Aspelmeyer_RMP_86_1391_2014}, we
rewrite the Hamiltonian \eqref{H_cav} in the frame rotating with the frequency $\omega_p$:
\begin{equation}\label{H_cav_RW}
  \hat{\mathcal{H}}
  = -\hbar(\delta + g\hat{x})(\hat{{\rm e}}^\dagger\hat{{\rm e}} + \sfrac{1}{2})
    + \hat{\mathcal{H}}_m + \hat{\mathcal{H}}_{\rm rest} \,,
\end{equation}
where $\delta=\omega_p-\omega_o$ is the detuning of the pump from cavity resonance. Then we extract explicitly from the field $\hat{{\rm e}}$ the classical mean part ${\rm E}$ created by the optical pump, $\hat{{\rm e}}\to{\rm E}+\hat{{\rm e}}$:
\begin{equation}
  \hat{\mathcal{H}}
  = -\hbar(\delta + g\hat{x})(
        |{\rm E}|^2  + {\rm E}^*\hat{{\rm e}} + {\rm E}\hat{{\rm e}}^\dagger
        + \hat{{\rm e}}^\dagger\hat{{\rm e}} + \sfrac{1}{2}
      )
    + \hat{\mathcal{H}}_m + \hat{\mathcal{H}}_{\rm rest} \,.
\end{equation}
The term $-\hbar(\delta + g\hat{x})|{\rm E}|^2$ here just create a static radiation pressure on the mechanical object, which can be compensated by some means; the term $-\hbar\delta({\rm E}^*\hat{{\rm e}} + {\rm E}\hat{{\rm e}}^\dagger)$ does not depend on $\hat{x}$ and we absorb it into $\mathcal{H}_{\rm rest}$; and the term $-\hbar g\hat{x}(\hat{{\rm e}}^\dagger\hat{{\rm e}} + \sfrac{1}{2})$ is of the second order of smallness and can be neglected. The remaining terms form the following canonical linearized optomechanical Hamiltonian:
\begin{equation}\label{H_cav_lin}
  \hat{\mathcal{H}}
  = -\hbar\delta(\hat{{\rm e}}^\dagger\hat{{\rm e}} + \sfrac{1}{2})
    - \hbar g({\rm E}^*\hat{{\rm e}} + \hc)\hat{x}
    + \hat{\mathcal{H}}_m + \hat{\mathcal{H}}_{\rm rest} \,.
\end{equation}

As the next step, consider the Michelson/Fabry-Perot interferometer shown in Fig.\,\ref{fig:dual_rec_Michelson}, {\it assuming the symmetry condition} (the consideration below actually reproduces in a simplified form the scaling law theorem of \cite{Buonanno2003}). Suppose here for simplicity that both recycling mirrors are absent. This scheme can be described by the sum of two single-mode Hamiltonians \eqref{H_cav} of the arm Fabry-Perot cavities:
\begin{equation}\label{H_MFP}
  \hat{\mathcal{H}}
  = \hbar\bigl[
        (\omega_o - g\hat{x}_N)(\hat{{\rm e}}_N^\dagger\hat{{\rm e}}_N + \sfrac{1}{2})
        + (\omega_o - g\hat{x}_E)(\hat{{\rm e}}_E^\dagger\hat{{\rm e}}_E + \sfrac{1}{2})
      \bigr]
    + \hat{\mathcal{H}}_m + \hat{\mathcal{H}}_{\rm rest} \,,
\end{equation}
where the subscripts $N$ and $E$ stand for the ``north'' and the ``east'' (as shown in Fig.\,\ref{fig:dual_rec_Michelson}) arms, respectively. This Hamiltonian, similar to \eqref{H_cav}, describes dispersive coupling.

Then introduce the common and the differential optical modes as follows:
\begin{align}\label{e_pm}
  \hat{{\rm e}}_\pm &= \frac{\hat{{\rm e}}_N \pm \hat{{\rm e}}_E}{\sqrt{2}} \,, &
  \hat{{\bf e}} &= \svector{\hat{{\rm e}}_+}{\hat{{\rm e}}_-} .
\end{align}
In these notations,
\begin{equation}\label{H_MFP_pm}
  \hat{\mathcal{H}}
  = \hbar\bigl[
        (\omega_o - g\hat{y})(\hat{{\bf e}}^\dagger\hat{{\bf e}} + 1)
        - g\hat{x}\hat{{\bf e}}^\dagger\mathbb{X}\hat{{\bf e}}
      \bigr]
    + \hat{\mathcal{H}}_m + \hat{\mathcal{H}}_{\rm rest} \,,
\end{equation}
where
\begin{align}
  y &= \frac{x_N + x_E}{2} \,, & x &= \frac{x_N - x_E}{2}
\end{align}
are coordinates of the common (symmetric) and the differential (antisymmetric) mechanical modes, and $\mathbb{X}$ is the Pauli $x$-matrix [see Eq.\,\eqref{2x2}]. For the common mode $y$, this Hamiltonian still retains the dispersive coupling structure. But the optomechanical coupling with the differential mode $x$ is of a different nature: in this case, the {\em coupling} of the two modes $\hat{{\rm e}}_+$ and $\hat{{\rm e}}_-$ is proportional to the mechanical displacement $x$. We will refer to this term as {\it coherent optomechanical coupling}. Note that opposite to \eqref{H_MFP}, the Hamiltonian \eqref{H_MFP_pm} is valid in the case of the general dual recycled interferometer as well \cite{Meers1988, Buonanno2003} and, in particular, in the case of the pure Michelson interferometer (without the ITM mirrors). In the particular case of a very broadband common optical mode, that is with the bandwidth much broader than all other characteristic frequencies of the system (with the evident exception of $\omega_o$, $\omega_p$), the common optical mode degenerates to an (almost) free space optical field. In this case, the bandwidth of the differential optical mode becomes dependent of $x$. This is the so-called {\it dissipative} optomechanical coupling \cite{Elste_PRL_102_207209_2009, Weiss_Burder_Nunnen_2012}. This simple example shows, that in multi-mode systems the type of the optomechanical coupling can not be categorized in a simple and unique way; it depends on a non-unique choice of the optical modes.

Now, following the above treatment of the Fabry-Perot cavity, we introduce explicitly the classical pumping fields by replacing $\hat{{\rm e}}_\pm \to {\rm E}_\pm + \hat{{\rm e}}_\pm$ and retrace the equations (\ref{H_cav_RW}-\ref{H_cav_lin}). This gives the following linearized Hamiltonian
\begin{equation}\label{H_MFP_pm_lin}
  \hat{\mathcal{H}} = -\hbar\delta
    (\hat{{\bf e}}^\dagger\hat{{\bf e}} + 1)
    - \hbar g\hat{y}({\bf E}^\dagger\hat{{\bf e}} + \hc)
    - \hbar g\hat{x}({\bf E}^\dagger\mathbb{X}\hat{{\bf e}} + \hc)
    + \hat{\mathcal{H}}_m + \hat{\mathcal{H}}_{\rm rest} \,,
\end{equation}
where
\begin{equation}
  {\bf E} = \svector{{\rm E}_+}{{\rm E}_-} .
\end{equation}
Note the similarity between this Hamiltonian and the one for the Fabry-Perot interferometer \eqref{H_cav_lin}.

Moreover, if the differential optical mode is not excited, ${\rm E}_-=0$ (which corresponds to the canonical regime of both the GW detectors and membrane interferometers), then the common optical mode is coupled only with the common mechanical one and the differential optical mode --- only with the differential mechanical one
\begin{equation}\label{H_MFP_pm_diag}
  \hat{\mathcal{H}} = -\hbar\delta
    (\hat{{\bf e}}^\dagger\hat{{\bf e}} + 1)
    - \hbar g\hat{y}({\rm E}_+^\dagger\hat{{\rm e}}_+ + \hc)
    - \hbar g\hat{x}({\rm E}_+^\dagger\hat{{\rm e}}_- + \hc)
    + \hat{\mathcal{H}}_m + \hat{\mathcal{H}}_{\rm rest} \,.
\end{equation}
Of these two mechanical modes, only the differential one is of interest in both the laser GW detectors and in the small-scale membrane interferometers. In the former case, it is this mode that is coupled with the gravitational waves. In the latter case, the mechanical common mode corresponds to the membrane thickness oscillations, which are characterized by very high (hundreds of gigahertz) eigenfrequency and low $Q$-factor and hardly can be used in optomechanical experiments. Therefore, the part of the Hamiltonian \eqref{H_MFP_pm_diag} referring to common modes can be omitted, which gives the following Hamiltonian
\begin{equation}\label{H_MFP_pm_diff}
  \hat{\mathcal{H}} = -\hbar\delta(\hat{{\rm e}}_-^\dagger\hat{{\rm e}}_- + \sfrac{1}{2})
    - \hbar g\hat{x}({\rm E}_+^\dagger\hat{{\rm e}}_- + \hc)
    + \hat{\mathcal{H}}_m + \hat{\mathcal{H}}_{\rm rest} \,.
\end{equation}
Up to the notations, it is identical to the Hamiltonian \eqref{H_MFP_pm_diff}, despite the completely different types of the optomechanical coupling --- the dispersive one in \eqref{H_cav_lin} and the coherent or the dissipative one in \eqref{H_MFP_pm_diff}.

\section{Analysis of the asymmetric interferometer}\label{sec:calculations}

Now, having discussed the various types of optomechanical coupling in Michelson-type interferometers, we are in position to consider in depth the asymmetric case. In the rest of this paper, we focus on the above mentioned case of a very broadband common optical mode, which is characterized by the dissipative (in contrast to coherent) optomechanical coupling. This case is typical for table-top interferometers researching fundamental optomechanics, because in this case much lower optical powers than in the large-scale gravitational-wave detectors is required. Due to this reason, we do not consider here the common mechanical mode. At the same time, both common and differential optical modes will be taken into account.

In the calculations below, we will use the Heisenberg picture (or input/output relations) approach which is more conveninent for analysis of sophisticated optomechanical systems, see \eg \cite{Caves1981, Buonanno2003, 12a1DaKh}. In this picture, the linearized dynamics of a two-port optomechanical system can be described by two matrix equations. The first one is the optical input/output relation:
\begin{equation}\label{in_out}
  \hat{{\bf b}}(\omega) = \mathbb{R}_{\rm ifo}(\omega)
    \bigl[\hat{{\bf a}}(\omega) + ik_p\mathbb{G}(\Omega){\bf E}\hat{x}(\Omega)\bigr]\,,
\end{equation}
where $\mathbb{R}_{\rm ifo}$ and $\mathbb{G}$ are $2\times2$ matrices and
\begin{align}
  \hat{{\bf a}} &= \svector{\hat{{\rm a}}_+}{\hat{{\rm a}}_-} , &
  \hat{{\bf b}} &= \svector{\hat{{\rm b}}_+}{\hat{{\rm b}}_-}
\end{align}
are two-components vectors for the input and output optical fields in the ``west'' and the ``south'' (as shown in Fig.\,\ref{fig:Michelson-Sagnac}) ports of the interferometer. The second equation describes the radiation pressure force acting on the mechanical object:
\begin{equation}\label{F}
  \hat{F}(\Omega) = \hat{F}_{\rm fl}(\Omega) - K(\Omega)\hat{x}(\Omega) \,,
\end{equation}
where
\begin{equation}\label{F_fl}
  \hat{F}_{\rm fl}(\Omega)
  = \hbar k_p{\bf E}^\dagger\mathbb{F}(\Omega)\hat{{\bf a}}(\omega) + \CC
\end{equation}
is the stochastic part of the radiation pressure force,
\begin{equation}\label{K}
  K(\Omega) = \hbar k_p^2{\bf E}^\dagger\mathbb{K}(\Omega){\bf E}
\end{equation}
is the optical rigidity, and $\mathbb{F}$, $\mathbb{K}$ are $2\times2$ matrices. The explicit equations for the matrices $\mathbb{R}_{\rm ifo}$, $\mathbb{G}$, $\mathbb{F}$, and $\mathbb{K}$ are quite cumbersome; they are derived in the Appendix, see Eqs.\,(\ref{bbR_ifo}, \ref{bbG}, \ref{bbF}, \ref{bbK}), respectively. The non-symmetrized spectral density $\tilde{S}_F$ of the force $\hat{F}_{\rm fl}$ can be obtained from Eq.\,\eqref{F_fl} using directly the definition \eqref{tilde_S}. In particular, if the incident quantum fields are in vacuum, then the spectral density is equal to
\begin{equation}\label{tilde_S_F}
  \tilde{S}_F(\Omega)
    = \hbar^2k_p^2{\bf E}^\dagger\mathbb{F}(\Omega)\mathbb{F}^\dagger(\Omega){\bf E}\,.
\end{equation}

An interesting feature of Eqs.\,\eqref{in_out} and \eqref{F} is the following symmetry condition [see Eqs.\,(\ref{bbG}, \ref{bbF})]:
\begin{equation}\label{G_F}
  \mathbb{G}(\Omega) = \mathbb{F}^\dagger(\Omega) \,.
\end{equation}
It is the two-port analog of the well-known relation between the measurement noise and the radiation pressure noise in ordinary (single-port) interferometers \cite{Caves1981, 02a1KiLeMaThVy, Buonanno2003}, which gives rise to the uncertainty relation between the radiation pressure noise and the measurement noise spectral densities of these devices \cite{Buonanno2003, 12a1DaKh} (which, in turn, is a particular case of the general uncertainty relation for the continuous linear quantum measurement \cite{92BookBrKh}).

As we have mentioned, in this paper we focus on the case without power-recycling,
\begin{equation}\label{no_pr}
  R_W=0 \,.
\end{equation}
In addition, we assume the lumped mode approximation (that is, the high finesse limit), which is a good approximation in common setups and significantly simplify the equations. Namely, we suppose that: (i) the transmissivity of the signal recycling mirror is small
\begin{subequations}\label{sm_app}
  \begin{gather}
    T_S^2 = 1-R_S^2 = 4\gamma_S\tau_S \ll 1 \,,
    \intertext{(ii) the signal recycling cavity is tuned close to the resonance:}
    e^{i\omega\tau_S} = e^{i(\delta_S+\Omega)\tau_S + i\theta} \,,\quad
      |\delta_S+\Omega|\tau_S \ll 1 \,, \\
    \intertext{where $\delta_S$ is the detuning of the ``south'' arm, and (iii) the asymmetry of the interferometer is small:}
      p^2 = \epsilon^2 + \varkappa^2 \ll 1 \,.
  \end{gather}
\end{subequations}
We assume the following relations between these small values:
\begin{equation}\label{nopr_sm_2}
  \gamma_S\tau_S \sim |\delta_S+\Omega|\tau_S \sim p^2 \,.
\end{equation}

Then, keeping {\it in each component} of the matrices $\mathbb{F}$ and $\mathbb{K}$ [see Eqs.\,(\ref{bbF}, \ref{bbK})] only the leading non-vanishing terms, we obtain that
\begin{gather}
  \mathbb{G}^\dagger(\Omega) = \mathbb{F}(\Omega) = \frac{2R_m}{\tau_S\ell(\Omega)}
    \smatrix{ip\sin(\alpha-\theta)}{\sqrt{\gamma_S\tau_S}\,e^{-i\theta}}
      {[\tau_S\ell_S(\Omega) + ip^2\sin2\alpha/2]e^{i\theta}}
      {-\sqrt{\gamma_S\tau_S}\,pe^{i(\theta-\alpha)}} , \label{bbF_nopr_sm} \\
  \mathbb{K}(\Omega) = -\frac{2iR_m}{\tau_S\ell(\Omega)}
    \smatrix{R_m}{-R_mpe^{-i\alpha}}{-R_mpe^{2i(\theta-\alpha)}}
      {[\tau_S\ell_S(\Omega) + \epsilon^2]e^{i\theta}}
  + \CC , \label{bbK_nopr_sm}
\end{gather}
where
\begin{subequations}
  \begin{gather}
    \ell(\Omega) = \gamma - i(\delta+\Omega) \,, \\
    \ell_S(\Omega) = \gamma_S - i(\delta_S+\Omega) \,,
  \end{gather}
\end{subequations}
\begin{subequations}
  \begin{gather}
    \gamma = \gamma_S + \gamma_m \,, \\
    \delta = \delta_S + \delta_m
  \end{gather}
\end{subequations}
are the total bandwidth and the detuning of the interferometer,
\begin{subequations}\label{gd_m}
  \begin{gather}
    \gamma_m = \frac{p^2\sin^2(\theta-\alpha)}{\tau_S} \,, \\
    \delta_m = \frac{p^2R_m\sin(\theta-2\alpha)}{\tau_S}
  \end{gather}
\end{subequations}
are the components of $\gamma$, $\delta$ due to the asymmetry  of the interferometer, and the angle $\alpha$ is defined as follows:
\begin{align}
  \epsilon &= p\cos\alpha \,, & \varkappa &= p\sin\alpha \,.
\end{align}
The dispersive and dissipative coupling factors can be readily derived from Eqs.\,\eqref{gd_m}:
\begin{subequations}\label{g_disp_diss}
  \begin{gather}
    g_{\rm disp} = -k_p\partd{\delta_m}{\varkappa}
      = \frac{2k_pR_mp}{\tau_S}\cos(\theta-\alpha)  \,, \\
    \frac{g_{\rm diss}}{\sqrt{2\gamma_m}} = k_p\partd{\sqrt{2\gamma_m}}{\varkappa}
      = \frac{2k_pR_m}{\sqrt{\tau_S}}\sign(\theta-\alpha)  \label{g_diss} 
  \end{gather}
\end{subequations}
(note that it is the combination \eqref{g_diss}, but not just $g_{\rm diss}$ appears in the dissipative coupling Hamiltonian, see \eg Eq.\,(1) of \cite{Xuereb_PRL_107_213604_2011}).

The upper row terms in the matrix \eqref{bbF_nopr_sm} has the order of magnitude of $\mathcal{O}(p^{-1})$, while the lower row ones --- of $\mathcal{O}(1)$. Correspondingly, the  matrix $\mathbb{F}\mathbb{F}^\dagger$, which appears in Eq.\,\eqref{tilde_S_F}, has the following structure:
\begin{equation}\label{smallnesses}
  \mathbb{F}(\Omega)\mathbb{F}^\dagger(\Omega) \sim
    \smatrix{\mathcal{O}(p^{-2})}{\mathcal{O}(p^{-1})}{\mathcal{O}(p^{-1})}{\mathcal{O}(1)}
    .
\end{equation}
Suppose now that either the classical field amplitudes ${\rm E}_\pm$ are of the same order of magnitude, or ${\rm E}_+$ dominates:
\begin{equation}\label{equalEs}
  {\rm E}_+ \gtrsim {\rm E}_- \,.
\end{equation}
In this case, the spectral density \eqref{tilde_S_F} is dominated by the term proportional to $|{\rm E}_+|^2$, with the other terms being small corrections which  have to be neglected for the sake of consistency with the already made approximations. This consideration gives the following equations for the non-symmetrized and symmetrized [see Eq.\,\eqref{S}] radiation pressure noise spectral densities:
\begin{subequations}\label{S_F_can}
  \begin{gather}
    \tilde{S}_F(\Omega)
      = \frac{4\hbar^2k_p^2R_m^2|{\rm E}_+|^2\gamma}{\tau_S|\ell(\Omega)|^2} \,, \\
    S_F(\Omega) = \frac{4\hbar^2k_p^2R_m^2|{\rm E}_+|^2\gamma}{\tau_S}\,
      \frac{\gamma^2+\delta^2+\Omega^2}{|\ell(\Omega)|^2|\ell(-\Omega)|^2} \,.
  \end{gather}
\end{subequations}
The matrix \eqref{bbK_nopr_sm} also has the structure \eqref{smallnesses}. Therefore, the above consideration is valid for the optical rigidity as well, giving:
\begin{equation}\label{K_can}
  K(\Omega)
  = \frac{4\hbar^2k_p^2R_m^2|{\rm E}_+|^2\delta}{\tau_S\ell(\Omega)\ell^*(-\Omega)} \,.
\end{equation}
Equations (\ref{S_F_can}, \ref{K_can}) do not depend on the interferometer asymmetry and differ from the well-know ``canonical'' ones \cite{Buonanno2003, 12a1DaKh} only by the well-expected factor $R_m^2$.

It follows from this consideration, that the ``non-canonical'' features, predicted in \cite{Xuereb_PRL_107_213604_2011, 13a1TaKhKaScHa} and observed in \cite{14a1SaKaNiTaKhHaSc}, evidently, originates from a violation of the assumption \eqref{equalEs}. In fact, it follows from Eq.\,\eqref{bf_E}, with account of the assumption \eqref{no_pr} and approximations \eqref{sm_app}, that the classical amplitudes of the intracavity fields are equal to
\begin{multline}
  {\bf E} = \frac{1}{\tau_S\ell(0)}
    \smatrix{\tau_S\ell_S(0) + ip^2\sin2\alpha/2}{-\!\sqrt{\gamma_S\tau_S}\,pe^{-i\alpha}}
      {ipe^{i\theta}\sin(\alpha-\theta)}{\sqrt{\gamma_S\tau_S}}
    \svector{{\rm A_W}e^{i\omega_p\tau_W}}{{\rm A_S}e^{i\omega_p\tau_S}} \\
  \sim\smatrix{\mathcal{O}(1)}{\mathcal{O}(1)}{\mathcal{O}(p^{-1})}{\mathcal{O}(p^{-1})}
    \svector{{\rm A_W}e^{i\omega_p\tau_W}}{{\rm A_S}e^{i\omega_p\tau_S}} ,
\end{multline}
which means that {\it typically}, instead of \eqref{equalEs},
\begin{equation}\label{differentEs}
  {\rm E}_- \sim \frac{{\rm E}_+}{p} \gg {\rm E}_+ \,.
\end{equation}
This resonance-enhanced value of ${\rm E}_-$ emphasizes the smaller terms in the matrices (\ref{bbF_nopr_sm}, \ref{bbK_nopr_sm}), making their contribution comparable with one of the ``canonical'' terms.

In particular, in the case of ${\rm A}_-=0$, which was considered in \cite{Xuereb_PRL_107_213604_2011, 13a1TaKhKaScHa},
\begin{equation}
  \tilde{S}_F(\Omega)
  = \frac{4\hbar^2k_p^2R_m^2|{\rm A}_+|^2}{\tau_S|\ell(0)|^2|\ell(\Omega)|^2}\Bigl\{
        \gamma_m(2\delta_S - 2\epsilon\varkappa/\tau_S + \Omega)^2
        + \gamma_S\bigl[\gamma^2 + (\delta_S-\delta_m-2\epsilon\varkappa/\tau_S)^2\bigr]
      \Bigr\} .
\end{equation}
Note the ``non-canonical'' Fano-resonance term, discussed in \cite{Elste_PRL_102_207209_2009, Xuereb_PRL_107_213604_2011}, which provides a minimum of $\tilde{S}_F(\Omega)$ at $\Omega=-2\delta_S+2\epsilon\varkappa/\tau_S$. It is evident, however, that by fine tuning of the values of ${\rm A}_{W,S}$, any ratio of ${\rm E}_+/{\rm E}_-$ can be obtained. In particular, as we show in the next section, the most effective optical cooling can be achieved by the ideally symmetric field, ${\rm E}_-=0$.

\section{Optimal optical cooling in Michelson-type interferometers}\label{sec:cooling}

In the recent experimental work \cite{14a1SaKaNiTaKhHaSc} optical cooling in the regime of interfering dispersive and dissipative coupling in an asymmetric Michelson-Sagnac interferometer was observed. Here we use our general framework to calculate the optimal cooling regime in the asymmetric Michelson-type interferometers for a given, fixed value of the optical power circulating in the interferometer.

We start with the two well known fundamental interrelations between any source of dissipation and the thermal noise $\hat{F}_T$ associated with it. The first one is the Fluctuation-Dissipation Theorem (FDT) \cite{Callen1951}:
\begin{equation}\label{FDT}
  S_T(\Omega) = \hbar|\Omega H|(2n_T + 1) \,,
\end{equation}
where
\begin{equation}
  S_T(\Omega) = \frac{\tilde{S}_T(\Omega) + \tilde{S}_T(-\Omega)}{2}
\end{equation}
is the symmetrized spectral density of this noise, $\tilde{S}_T(\Omega)$ is the corresponding non-symmetrized spectral density, see Eqs.\,(\ref{tilde_S}, \ref{S}), $H$ is the friction factor,  $n_T$ is the effective number of thermal quanta defined by
\begin{equation}
  2n_T + 1 = \coth\frac{\hbar|\Omega|}{2\kappa_BT} \,,
\end{equation}
and $T$ is the temperature.  The second one is the Kubo theorem \cite{Kubo1956}:
\begin{equation}\label{Kubo}
  \Omega H = \frac{\tilde{S}_T(\Omega) - \tilde{S}_T(-\Omega)}{2\hbar} \,.
\end{equation}
Assuming that $H>0$ (stable system dynamics) and $\Omega>0$, it is easy to get from Eqs.\,(\ref{FDT}, \ref{Kubo}), that
\begin{equation}
  \frac{1}{n_T} +1 = \frac{\tilde{S}_T(\Omega)}{\tilde{S}_T(-\Omega)} \,.
\end{equation}

In optical cooling experiments, the ``native'' mechanical heat bath is supplemented by the low temperature optomechanical one. In this case, the steady state mean number of phonons in the mechanical oscillator is given by
\begin{equation}
  2\mean{n} + 1 = \frac{S_T(\Omega_m) + S_F(\Omega_m)}{\hbar\Omega_m(H + H_{\rm opt})} \,,
\end{equation}
where $S_F$ is the symmetrized spectral density of the radiation pressure noise, $H_{\rm opt}$ is the optical damping:
\begin{equation}
  \Omega H_{\rm opt} = -\Im K \,,
\end{equation}
and we absorbed the shift of the mechanical resonance frequency imposed by the optical spring into $\Omega_m$.

Rewriting the Kubo theorem for the optical damping:
\begin{equation}
  \Omega H_{\rm opt} = \frac{\tilde{S}_F(\Omega) - \tilde{S}_F(-\Omega)}{2\hbar} \,,
\end{equation}
it is is easy to show that
\begin{equation}\label{mean_n}
  \frac{1}{\mean{n}} + 1 = \frac{\tilde{S}_T(\Omega_m) + \tilde{S}_F(\Omega_m)}
    {\tilde{S}_T(-\Omega_m) + \tilde{S}_F(-\Omega_m)} \,.
\end{equation}

In the Michelson-Sagnac interferometer, the explicit form of $\tilde{S}_F$ is rather sophisticated, see Eqs.\,(\ref{tilde_S_F}, \ref{bbF_nopr_sm}), and the direct analytical optimization of \eqref{mean_n} is hardly possible. However, under common experimental conditions the spectral densities $\tilde{S}_F$ and $\tilde{S}_T$ satisfy strong inequalities which significantly simplify this task. Really, starting values of the thermal occupation number of the real mechanical resonators are big, even in the cryogenic microwave experiments; correspondingly,  asymmetry of the thermal noise spectral density is small:
\begin{equation}
  \tilde{S}_T(\Omega_m) - \tilde{S}_T(-\Omega_m) \ll \tilde{S}_T(\pm\Omega_m) \,.
\end{equation} 
Therefore, in order to provide effective optical cooling, asymmetry of the radiation pressure noise spectral density has to be strong:
\begin{equation}
  \tilde{S}_F(\Omega_m) \gg \tilde{S}_F(-\Omega_m) \,.
\end{equation}
At the same time, due to technical constrains in contemporary optical cooling experiments, while $\tilde{S}_F(\Omega_m)$ could be close or even exceeds the thermal noise spectral density, its negative-frequency counterpart is small:
\begin{equation}
  \tilde{S}_F(-\Omega_m) \ll \tilde{S}_T(\pm\Omega_m) \,.
\end{equation}
(this inequality was fulfilled with very good margin even in the record-breaking works \cite{Teufel_Nature_475_359_2011, Chan_Nature_478_89_2011}). These assumptions simplify Eq.\,\eqref{mean_n} to
\begin{equation}\label{mean_n_opt}
  \mean{n} = \frac{\tilde{S}_T(-\Omega_m)}{\tilde{S}_F(\Omega_m)} \,.
\end{equation}
In this case, minimization of $\mean{n}$ is simply equivalent to maximization of $\tilde{S}_F(\Omega)$.

Of the mentioned above technical constrains, the most serious ones are limitations on the value of the optical power inside the interferometer imposed by various undesirable effects, like heating, mechanical nonlinearities, instabilities \etc. Therefore consider maximization of $\tilde{S}_F(\Omega)$, assuming a given optical energy in the interferometer, which is proportional to
\begin{equation}\label{sum_E}
  \mathcal{E} \propto |{\rm E}_+|^2 + |{\rm E}_-|^2 \,.
\end{equation}
It follows form Eqs.\,(\ref{tilde_S_F}, \ref{smallnesses}), that this spectral density has the following structure:
\begin{equation}
  \tilde{S}_F \propto \mathcal{O}(p^{-2})|{\rm E}_+|^2
  + 2\mathcal{O}(p^{-1})\Re({\rm E}_+^*{\rm E}_-)  + \mathcal{O}(1)|{\rm E}_-|^2  \,,
\end{equation}
that is, the symmetric field ${\rm E}_+$ provides the largest value of $\tilde{S}_F$ and therefore the most effective cooling. Therefore, with account of the optical energy constrain, the antisymmetric field has to be canceled, ${\rm E}_-=0$. In this case, the radiation pressure noise spectral density reduces to the canonical Lorentzian form \eqref{S_F_can}. 

\section{Summary}\label{sec:summary}

We have shown that the standard description of the radiation-pressure induced optomechanical coupling as either ``dispersive'' or ``dissipative'' is univocal only in the simplest case of a single lumped electromagnetic mode. In the general multi-mode case, in particular in Michelson-type interferometers, the coupling type depends on the non-unique choice of its optical modes.

The most convenient choice, broadly used by the GW community, uses the common and differential optical modes of the interferometer, where the differential optical mode couples to the conventional signal output port. For these modes, the type of the optomechanical coupling further depends on whether the power recycling technique (in addition to signal-recycling) is used or not. In the latter case, the coupling is dissipative, with a dispersive contribution if the interferometer is not perfectly \emph{symmetric}. In the former one, a more sophisticated behavior emerges, where the coupling between two optical modes depends on the mechanical displacement, which we coined as the ``coherent optomechanical coupling''.

We have developed a general framework to calculate the optomechanical properties of the Michelson-type interferometers in the \emph{asymmetric} regime. It covers the possibility of the injection of carrier light into both ports of the interferometer. We used this framework for in depth analysis of the radiation pressure features (both dynamic and stochastic) of the Michelson-type interferometers without the power recycling, leaving the power-recycled configuration, with its different modes and optomechanical coupling structure, for future work.

Our analysis has shown that the ``anomalous'' features originate from the small second-order terms in the Taylor expansion of the (non-symmetrized) radiation pressure noise spectral density in the interferometer length and its asymmetry, see Eqs.\,(\ref{sm_app}, \ref{nopr_sm_2}). Usually, these terms are ignored in the lumped modes approximation routinely used in the analysis in the quantum optomechanical setups. In unbalanced Michelson-type interferometers these corrections are strongly amplified by the resonance-enhanced optical power in the differential optical mode of the interferometer and therefore change significantly the interferometer behavior.

Finally, we have shown that under common experimental conditions, and for a given optical power {\em inside} a cavity-enhanced Michelson interferometer, the lowest steady state mean phonons number $\mean{n}$ can be achieved by exciting the common optical mode alone with balanced light power in both arms. In this case the operation regime of the interferometer is ``canonical'' and fully corresponds to optical cooling in a Fabry-Perot cavity with dispersive coupling. At the same time, both dispersive and dissipative type of coupling could coexist in this case (however, optimal cooling regimes for the two-mode dual recycled interferometers and/or for a given \emph{injected} light power, could differ from this).

\acknowledgments

This work was supported by the Marie Curie Initial Training Network cQOM, by the ERC Advanced Grant MassQ, by the International Max Planck Research School for Gravitational Wave Astronomy (IMPRS), and by DFG through Research Training Group 1991 {\em Quantum mechanical noise in complex systems}. The work of F. Khalili was supported by LIGO NSF Grant No\,PHY-1305863 and Russian Foundation for Basic Research Grant No.\,14-02-00399.

The authors thank Haixing Miao for useful remarks.

The paper has been assigned LIGO document number P1600065.

\appendix

\begin{figure}
  \includegraphics{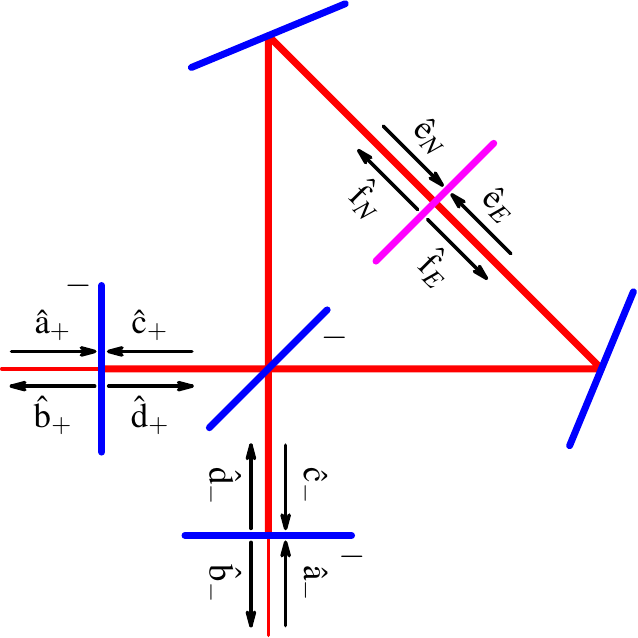}
  \caption{To calculation of optical fields in the Michelson-Sagnac interferometer.}\label{fig:the_scheme_app}
\end{figure}

\section{Notations}\label{app:notations}

\begin{align}\label{2x2} 
  \mathbb{I} &= \smatrix{1}{0}{0}{1} \,, & \mathbb{Z} &= \smatrix{1}{0}{0}{-1} \,, &
  \mathbb{X} &= \smatrix{0}{1}{1}{0} \,, & \mathbb{Y} &= \smatrix{0}{-1}{1}{0} 
\end{align}
are the general-purposes $2\times2$ matrices.

We describe the membrane by the symmetric reflectivity/transmissivity matrix
\begin{equation}
  \smatrix{R_m}{iT_m}{iT_m}{R_m} ,
\end{equation}
and the beamsplitter and the recycling mirrors --- by real ones
\begin{align}
  &\smatrix{R}{T}{T}{-R} , &&\smatrix{R_{W,S}}{T_{W,S}}{T_{W,S}}{-R_{W,S}} ,
\end{align}
with the negative reflectivities indicated by ``$-$'' in Fig.\,\ref{fig:the_scheme_app}. The quantum field sideband amplitudes are denoted by the lowercase roman letters
\begin{equation}
  \hat{{\rm a}}\,, \hat{{\rm b}}\,, \dots
\end{equation}
and the classical amplitudes --- by the corresponding uppercase roman ones
\begin{equation}
  {\rm A}\,, {\rm B}\,, \dots.
\end{equation}
The non-symmetrized spectral density $\tilde{S}$ of any noise process $\hat{F}$ is defined by
\begin{equation}\label{tilde_S}
  \Mean{\hat{F}(\Omega)\hat{F}(\Omega')} = 2\pi\tilde{S}(\Omega)\delta(\Omega+\Omega') \,,
\end{equation}
and the corresponding symmetrized one $S$ --- by
\begin{equation}\label{S}
  S(\Omega) = \frac{\tilde{S}(\Omega) + \tilde{S}(-\Omega)}{2} \,.
\end{equation}

\section{Optical fields}

We  assume that the D.C. displacement $X$ is small and neglect the term $(k-k_p)X = \Omega X/c$. In this case, the equations for the quantum field sideband amplitudes are the following \cite{12a1DaKh} [see the notations in Fig.\,\eqref{fig:the_scheme_app}]:
\begin{subequations}\label{eqs_1}
  \begin{align}
    \hat{{\rm b}}_+(\omega)
      &=-R_W\hat{{\rm a}}_+(\omega) + T_W\hat{{\rm c}}_+(\omega) \,, \\
    \hat{{\rm b}}_-(\omega)
      &=-R_S\hat{{\rm a}}_-(\omega) + T_S\hat{{\rm c}}_-(\omega) , \\
    \hat{{\rm c}}_+(\omega) &= \bigl[
        R\hat{{\rm f}}_N(\omega)e^{i\varkappa} + T\hat{{\rm f}}_E(\omega)e^{-i\varkappa}
      \bigr]e^{i\omega\tau_W} \,, \\
    \hat{{\rm c}}_-(\omega) &= \bigl[
        T\hat{{\rm f}}_N(\omega)e^{i\varkappa} - R\hat{{\rm f}}_E(\omega)e^{-i\varkappa}
      \bigr]e^{i\omega\tau_S} \,, \\
    \hat{{\rm d}}_+(\omega)
      &=T_W\hat{{\rm a}}_+(\omega) + R_W\hat{{\rm c}}_+(\omega) \,, \\
    \hat{{\rm d}}_-(\omega)
      &=T_S\hat{{\rm a}}_-(\omega) + R_S\hat{{\rm c}}_-(\omega) , \\
    \hat{{\rm e}}_N(\omega) &= \bigl[
        R\hat{{\rm d}}_+(\omega)e^{i\omega\tau_W}
        + T\hat{{\rm d}}_-(\omega)e^{i\omega\tau_S}
      \bigr]e^{i\varkappa} \,, \\
    \hat{{\rm e}}_E(\omega) &= \bigl[
        T\hat{{\rm d}}_+(\omega)e^{i\omega\tau_W}
        - R\hat{{\rm d}}_-(\omega)e^{i\omega\tau_S}
      \bigr]e^{-i\varkappa}  \,, \\
    \hat{{\rm f}}_N(\omega) &= R_m\hat{{\rm e}}_N(\omega) + iT_m\hat{{\rm e}}_E(\omega)
      + 2ik_pR_m{\rm E}_N\hat{x}(\Omega) \,, \\
    \hat{{\rm f}}_E(\omega) &= R_m\hat{{\rm e}}_E(\omega) + iT_m\hat{{\rm e}}_N(\omega)
      - 2ik_pR_m{\rm E}_E\hat{x}(\Omega) \,.
  \end{align}
\end{subequations}
Introduce the common and differential optical modes:
\begin{align}
  \hat{{\rm e}}_\pm &= \frac{\hat{{\rm e}}_N \pm \hat{{\rm e}}_E}{\sqrt{2}} \,, &
  \hat{{\rm f}}_\pm &= \frac{\hat{{\rm f}}_N \pm \hat{{\rm f}}_E}{\sqrt{2}} \,.
\end{align}
Using these modes:
\begin{subequations}\label{eqs_1_1}
  \begin{align}
    \hat{{\bf b}}(\omega)
      &=-\mathbb{R}\hat{{\bf a}}(\omega) + \mathbb{T}\hat{{\bf c}}(\omega) \,,
      \label{eqs_1_1(b)} \\
    \hat{{\bf c}}(\omega) &= \mathbb{A}(\omega)\mathbb{Q}^{\sf T}\hat{{\bf f}}(\omega)
      \label{eqs_1_1(c)} \,, \\
    \hat{{\bf d}}(\omega)
      &=\mathbb{T}\hat{{\bf a}}(\omega) + \mathbb{R}\hat{{\bf c}}(\omega) \,, \\
    \hat{{\bf e}}(\omega) &= \mathbb{Q}\mathbb{A}(\omega)\hat{{\bf d}}(\omega)\,,
      \\
    \hat{{\bf f}}(\omega) &= \mathbb{M}\hat{{\bf e}}(\omega)
      + 2ik_pr\mathbb{X}{\bf E}\hat{x}(\Omega)\,, \label{eqs_1_1(f)}
  \end{align}
\end{subequations}
where
\begin{subequations}
  \begin{align}
    \hat{{\bf a}} &= \svector{\hat{{\rm a}}_+}{\hat{{\rm a}}_-} , &
    \hat{{\bf b}} &= \svector{\hat{{\rm b}}_+}{\hat{{\rm b}}_-} , \\
    \hat{{\bf c}} &= \svector{\hat{{\rm c}}_+}{\hat{{\rm c}}_-} , &
    \hat{{\bf d}} &= \svector{\hat{{\rm d}}_+}{\hat{{\rm d}}_-} , \\
    \hat{{\bf e}} &= \svector{\hat{{\rm e}}_+}{\hat{{\rm e}}_-} , &
    \hat{{\bf f}} &= \svector{\hat{{\rm f}}_+}{\hat{{\rm f}}_-} ,
  \end{align}
\end{subequations}
\begin{gather}
  \mathbb{R} = \smatrix{R_W}{0}{0}{R_S} , \qquad
  \mathbb{T} = \smatrix{T_W}{0}{0}{T_S} , \\
  \mathbb{A}(\omega) = \smatrix{e^{i\omega\tau_W}}{0}{0}{e^{i\omega\tau_S}} , \\
  \mathbb{M} = \smatrix{e^{i\theta}}{0}{0}{e^{-i\theta}} , \\
  \mathbb{Q} = \smatrix{C}{-S^*}{S}{C^*} ,
\end{gather}
\begin{subequations}
  \begin{align}
    C &= \cos\epsilon\cos\varkappa + i\sin\epsilon\sin\varkappa \,,\\
    S &= \sin\epsilon\cos\varkappa + i\cos\epsilon\sin\varkappa \,.
  \end{align}
\end{subequations}

Eqs.\,\eqref{eqs_1_1} can be reduced to the following two:
\begin{subequations}\label{eqs_1_3}
  \begin{align}
    \mathbb{D}_e(\omega)\hat{{\bf e}}(\omega)
      &=\tilde{\mathbb{T}}(\omega)\hat{{\bf a}}(\omega)
        + 2ik_pr\tilde{\mathbb{R}}(\omega)\mathbb{Q}^{\sf T}\mathbb{X}{\bf E}x(\Omega)\,,
        \\
    \mathbb{D}_e(\omega)\mathbb{M}^\dagger\hat{{\bf f}}(\omega)
      &=\tilde{\mathbb{T}}(\omega)\hat{{\bf a}}(\omega)
        + 2ik_pr\mathbb{Q}^\dagger\mathbb{M}^\dagger\mathbb{X}{\bf E}x(\Omega)\,,
  \end{align}
\end{subequations}
where
\begin{gather}
  \mathbb{D}_e(\omega)
    = \mathbb{Q}^\dagger - \tilde{\mathbb{R}}(\omega)\mathbb{Q}^{\sf T}\mathbb{M} \,,
    \label{D_1}  \\
  \tilde{\mathbb{R}}(\omega) = \mathbb{A}(\omega)\mathbb{R}\mathbb{A}(\omega)
    = \smatrix{\tilde{R}_W(\omega)}{0}{0}{\tilde{R}_S(\omega)} , \\
  \tilde{\mathbb{T}}(\omega) = \mathbb{A}(\omega)\mathbb{T}
    = \smatrix{\tilde{T}_W(\omega)}{0}{0}{\tilde{T}_S(\omega)} ,
\end{gather}
\begin{align}
  \tilde{R}_{W,S}(\omega) &= R_{W,S}e^{2i\omega\tau_{W,S}}\,, &
  \tilde{T}_{W,S}(\omega) &= T_{W,S}e^{i\omega\tau_{W,S}}\,.
\end{align}
The solution to Eqs.\,\eqref{eqs_1_3} is
\begin{subequations}\label{eqs_1_soln}
  \begin{align}
    \hat{{\bf e}}(\omega) &= \mathbb{D}_e^{-1}(\omega)\bigl[
        \tilde{\mathbb{T}}(\omega)\hat{{\bf a}}(\omega)
        + 2ik_pr\tilde{\mathbb{R}}(\omega)\mathbb{Q}^{\sf T}\mathbb{X}{\bf E}x(\Omega)
      \bigr] \label{eqs_1_soln(e)}, \\
    \hat{{\bf f}}(\omega) &= \mathbb{M}\mathbb{D}_e^{-1}(\omega)\bigl[
        \tilde{\mathbb{T}}(\omega)\hat{{\bf a}}(\omega)
        + 2ik_pr\mathbb{Q}^\dagger\mathbb{M}^\dagger\mathbb{X}{\bf E}x(\Omega)
      \bigr] , \label{eqs_1_soln(f)}
  \end{align}
\end{subequations}
where
\begin{gather}
  \mathbb{D}_e^{-1}(\omega)
    = \frac{\mathbb{Q} - \mathbb{M}^\dagger\mathbb{Q}^*\breve{\mathbb{R}}(\omega)}
        {D(\omega)}\,,  \label{inv_D_1}  \\
  \breve{\mathbb{R}}(\omega) = \smatrix{\tilde{R}_S(\omega)}{0}{0}{\tilde{R}_W(\omega)},\\
  D(\omega) = \det\mathbb{D}_e(\omega) \,.
\end{gather}

Then, it follows from Eqs.\,(\ref{eqs_1_1(b)}, \ref{eqs_1_1(c)}, \ref{eqs_1_soln(f)}), that:
\begin{equation}
  \hat{{\bf b}}(\omega) =-\mathbb{R}\hat{{\bf a}}(\omega)
    + \mathbb{\tilde{T}}(\omega)\mathbb{Q}^{\sf T}\hat{{\bf f}}(\omega) \,,
\end{equation}
which gives Eq.\,\eqref{in_out} with
\begin{gather}
  \mathbb{R}_{\rm ifo}(\omega) = -\mathbb{R} + \frac{
      \tilde{\mathbb{T}}(\omega)
        \bigl[\mathbb{Q}^{\sf T}\mathbb{M}\mathbb{Q} - \breve{\mathbb{R}}(\omega)\bigr]
        \tilde{\mathbb{T}}(\omega)
    }{D(\omega)} \,, \label{bbR_ifo} \\
  \mathbb{G}(\Omega) = \frac{2r}{D^*(\Omega)}\mathbb{T}^\dagger(\Omega)\bigl[
        \mathbb{Q}^\dagger\mathbb{M}^\dagger
        - \breve{\mathbb{R}}^\dagger(\Omega)\mathbb{Q}^{\sf T} 
      \bigr]\mathbb{X}     
    \,. \label{bbG}
\end{gather}

The classical amplitudes vector ${\bf E}$ can be obtained from Eq.\,\eqref{eqs_1_soln(e)} by setting there $\omega=\omega_p$ and $x=0$:
\begin{equation}\label{bf_E}
  {\bf E} = \mathbb{D}_e^{-1}(\omega_p)\tilde{\mathbb{T}}(\omega_p){\bf A} \,.
\end{equation}

\section{Radiation pressure force}

The A.C. optical force acting on the membrane is equal to
\begin{multline}
  \hat{F}(\Omega) = \hbar k_p\bigl[
      {\rm E}_N^*\hat{{\rm e}}_N(\omega_p+\Omega) +
      {\rm F}_N^*\hat{{\rm f}}_N(\omega_p+\Omega)
      - {\rm E}_E^*\hat{{\rm e}}_E(\omega_p+\Omega)
      - {\rm F}_E^*\hat{{\rm f}}_E(\omega_p+\Omega)
    \bigr] + \CC \\
  = \hbar k_p\bigl[
        {\bf E}^\dagger\mathbb{X}\hat{{\bf e}}(\omega_p+\Omega)
        + {\bf F}^\dagger\mathbb{X}\hat{{\bf f}}(\omega_p+\Omega)
      \bigr] + \CC ,
\end{multline}
where
\begin{equation}\label{def_CC}
  \forall f(\omega): f(\Omega) + \CC = f(\Omega) + f^\dagger(-\Omega)
\end{equation}
and the dagger means the Hermitian conjugation both for matrices and quantum operators.

It follows from Eq.\,\eqref{eqs_1_1(f)}, that
\begin{equation}
  \hat{F}(\Omega) = \hat{F}_1(\Omega) - K_2(\Omega)x(\Omega) \,,
\end{equation}
where
\begin{equation}
  \hat{F}_1(\Omega) = 2\hbar k_pr{\bf E}^\dagger\mathbb{X}\mathbb{M}\hat{{\bf e}}(\Omega)
\end{equation}
and
\begin{subequations}
  \begin{gather}
    K_2(\Omega) = \hbar k_p^2{\bf E}^\dagger\mathbb{K}_2(\Omega){\bf E} \,, \\
    \mathbb{K}_2(\Omega) = -2ir\mathbb{M}^\dagger + \CC = -4rt\mathbb{Z}
  \end{gather}
\end{subequations}
is the part of the optical rigidity created by electrostatic attraction of the membrane into the standing wave antinode.

Then, using Eq.\,\eqref{eqs_1_soln}, we obtain, that:
\begin{equation}
  \hat{F}_1(\Omega) = \hat{F}_{\rm fl}(\Omega) - K_1(\Omega)x(\Omega) \,,
\end{equation}
where $\hat{F}_{\rm fl}$ is the stochastic force described by Eq.\,\eqref{F_fl}, with
\begin{equation}
  \mathbb{F}(\Omega) = \frac{2r}{D(\Omega)}\mathbb{X}
    \bigl[\mathbb{M}\mathbb{Q} - \mathbb{Q}^*\breve{\mathbb{R}}(\omega)\bigr]
    \tilde{\mathbb{T}}(\omega) \,, \label{bbF}
\end{equation}
and
\begin{subequations}
  \begin{gather}
    K_1(\Omega) = \hbar k_p^2{\bf E}^\dagger\mathbb{K}_1(\Omega){\bf E} \,, \\
    \mathbb{K}_1(\Omega) = -\frac{4ir^2}{D(\Omega)}\mathbb{X}\bigl[
        \mathbb{M}\mathbb{Q}\tilde{\mathbb{R}}(\omega)\mathbb{Q}^{\sf T}
        - \tilde{R}_W(\omega)\tilde{R}_S(\omega)\mathbb{I}
      \bigr]\mathbb{X} + \CC
  \end{gather}
\end{subequations}
is the part of the optical rigidity created by the modulation of the intracivity optical field by the membrane motion, that is the optical spring proper.

Correspondingly, the total optical rigidity
\begin{equation}
  K(\Omega) = K_1(\Omega) + K_2(\Omega)
\end{equation}
is equal to \eqref{K}, with
\begin{equation}\label{bbK}
  \mathbb{K}(\Omega) = \mathbb{K}_1(\Omega) + \mathbb{K}_2(\Omega) \,.
\end{equation}


\end{document}